\documentstyle[twocolumn,prb,aps,psfig]{revtex}
\def\be{\begin{equation}}
\def\ee{\end{equation}}
\def\bc{\begin{center}}
\def\ec{\end{center}}

\begin{document}
 
\input epsf.sty
 
\draft
 
\title{Microcanonical determination of the order parameter critical exponent}
 
\author{Alfred H\"{u}ller and Michel Pleimling}
 
\address{
Institut f\"ur Theoretische Physik I, Universit\"at Erlangen-N\"urnberg,
D -- 91058 Erlangen, Germany}
\maketitle

\begin{abstract}
A highly efficient Monte Carlo method for the calculation
of the density of states of classical spin systems is presented.
As an application, 
we investigate the density of states $\Omega_N(E,M)$
of two- and three-dimensional Ising models with $N$ spins as a function 
of energy $E$ and magnetization $M$. For a fixed energy lower than a 
critical value $E_{c,N}$ the density of states exhibits two sharp maxima at
$M = \pm M_{sp}(E)$ which define the microcanonical spontaneous magnetization. 
An analysis
of the form $M_{sp}(E) \propto (E_{c,\infty}-E)^{\beta_\varepsilon}$ yields very 
good results for the critical exponent $\beta_\varepsilon$,
thus demonstrating that critical exponents can be determined by analysing directly
the density of states of finite systems.
\end{abstract}
 
 
 
\section{Introduction}
Substantial progress has been achieved in the understanding of
discontinuous phase transitions since the problem has been formulated
in microcanonical terms\cite{Gro,Hue94}. Till now the microcanonical analysis of
continuous phase transitions has only been partly successful. On the
one hand it is a great accomplishment that the typical features of
symmetry breaking, i.e. the abrupt onset of the order parameter as the
critical point is approached from above and the diverging
susceptibilities, turn up already for finite systems\cite{Kas00}. This is
in contrast to the canonical ensemble where singularities appear
exclusively in the thermodynamic limit\cite{Bar83}. On the other hand it is an
intriguing and disappointing fact that mean field values have been
found for all critical exponents and for all finite system sizes\cite{Kas00}
when directly analysing the density of states.
 
We do not contest the earlier analysis where values of
the critical exponents of an Ising system were determined from the
expansion of the entropy $S_N(E,M)$ in the vicinity of the critical
point of the finite system, situated at $E = E_{N,c}$ and $M = 0$. Here it is our
aim to show that the non classical infinite lattice exponents can be
obtained directly from the density of states
if the expansion of the finite system entropy $S_N(E,M)$ is
carried out at the critical point $E = E_{c,\infty}$ and $ M = 0$ of the
infinite system. (The critical point of the finite or infinite system
is defined as the value of the energy $E = E_{c,N}$, where the second
derivative $[\partial ^2S_N/\partial M^2]_{E}$ of the entropy taken at $M=0$
changes its sign.)
 
For the purpose of determining the density of states $\Omega_N(E,M) =
\exp S_N(E,M)$ (in units where $k_B=1$) we have developed a highly efficient algorithm
which is based upon the Broad Histogram method\cite{Oli96,Oli98}. It allows a precise
and speedy determination of the density of states as a function of
one, two or even more parameters should it be necessary.
In the course of these calculations it will be
demonstrated that very good values for the order parameter critical exponent
can be obtained by a microcanonical analysis already for relatively small
system sizes.

The paper is organized in the following way. In the next Section we present
the numerical method which permits us to compute the density of states of
classical spin systems with very high accuracy. Section III is devoted to the
analysis of the microcanonically defined spontaneous magnetization and to the
determination of the corresponding critical exponent. A short summary concludes
the paper. 

\section{Numerical method}
One of the aims of computer simulations in statistical mechanics is
the determination of averages of observables $\hat{X}$
of the many particle system under study:

\begin{equation} \label{Gl:1}
\left< \hat{X} \right>_{\hat W} = \sum\limits_{\mu \in \Gamma_\mu} \hat{W} ( \mu ) \, 
\hat{X} ( \mu ).
\end{equation}
The space $\Gamma_\mu$ contains an enormous number $\Lambda_\mu$ of
microstates and 
the sum in (\ref{Gl:1}) runs over all of them.
$\hat{W}(\mu)$ is a probability
distribution on that space and $\hat{X}(\mu)$ is the value of
$\hat{X}$ in the microstate $\mu$.

In all cases of interest $\hat{W}(\mu)$ and $\hat{X}(\mu)$ depend on
$\mu$ only via a few parameters $\underline{P} = (P_a,P_b,...)$, i.e.,

\begin{equation} \label{Gl:2} 
\hat{W}(\mu) = W(\underline{P}(\mu))
\end{equation}
and

\begin{equation} \label{Gl:3}
\hat{X}(\mu) = X(\underline{P}(\mu))
\end{equation} 
Examples for $\underline{P}$ might be the interaction energy $E$ of a
magnetic system and its magnetization $M$. An example for
$W(\underline{P})$ could be the canonical distribution

\begin{equation} \label{Gl:4}
W(E,M)=\frac{\exp \left[ - \beta \left(E - h \, M \right) \right]}
{\sum\limits_{E',M'} \exp \left[ - \beta \left(E' - h \, M' \right) \right]}.
\end{equation}
The parameters $\underline{P}$ characterize the macrostate of the
system and examples for $X(\underline{P})$ might be the parameters
themselves, their fluctuations as e.g. $ (E - <E>_W)^2 $ or any other function
of them.

The unwieldy sum in (\ref{Gl:1}) is greatly reduced if the
degeneracies $\Omega(\underline{P})$ of the macrostates which are
characterized by the parameters $\underline{P}$ are known. Then:

\begin{equation} \label{Gl:5}
\left< \hat{X} \right>_{\hat W} = \left< X \right>_W = \sum\limits_{\underline{P}}
 \Omega(\underline{P}) \, W(\underline{P}) \, X(\underline{P}).
\end{equation}
$\Omega(\underline{P})$ is also called the density of states DOS. The
number of terms in the sum (\ref{Gl:1}) is already gigantesque for a
system of only a few hundred particles: $\Lambda_\mu$ is of order $f^N$
where $N$ is the number of particles in the system. $f$ is equal to 2
for an Ising system and for a $q$-states Potts model $f = q$.

Compared to that, the sum in (\ref{Gl:5}) can be performed with very
little effort indeed. Its number of terms is only of order $N^\nu$
where $\nu$ is the number of parameters. Once
$\Omega(\underline{P})$ is known, the sum in (\ref{Gl:5}) can be
performed in the twinkle of an eye by any desktop computer.

Now the difficulty of calculating averages of $X$ has been shifted to
the determination of the DOS. This does not seem to be any easier, but
we have developed a new algorithm to construct
$\Omega(\underline{P})$. 
Before the algorithm is explained in detail we
want to point at yet another advantage which is offered at no extra
cost when the density of states $\Omega(E,M)$ has been determined. As
$\ln \Omega(E,M) = S(E,M)$ (setting $k_B=1$) is the entropy as a function of its natural
variables, the thermodynamic relations can be derived directly from
$S(E,M)$ by differentiation. One thereby obtains the microcanonically
defined temperature, specific heat, susceptibilities etc.

This is an extra benefit on top of the possibility of having the mean
values of any function $X(\underline{P})$ (e.g.\ $X(E,M)$) and for any
desired probability distribution $W(\underline{P})$ at hand.

The algorithm is a variant of the transition variable method\cite{Oli96,Oli98,Kasb00}. Its
starting point is an extremely simple observation: Consider any
mechanism which generates a new microstate $\mu'$ when it is applied
on the microstate $\mu$. An example for such a mechanism could be a
single spin flip in an Ising system. When the single spin flip mechanism is applied to
all the $N$ spins of all the $\Omega(\underline{P}_i)$ microstates 
which belong to the macrostate which is
characterized by $\underline{P}_i$,
then $N \, \Omega(\underline{P}_i)$ new microstates are created. Let $V_{ij}$ be the number
of them which belong to the macrostate characterized by $\underline{P}_j$. As any spin
flip can be reversed: $V_{ji} = V_{ij}$. In other words there are as many connections
(by single spin flip) from level $\underline{P}_i$ to level $\underline{P}_j$ as there are
connections from level $\underline{P}_j$ to $\underline{P}_i$. Thus if one chooses one
microstate of level $\underline{P}_i$ at random, selects one spin at random, and flips it,
the probability of arriving at level $\underline{P}_j$ will be $V_{ij} / (N \, \Omega(\underline{P}_i))$
and vice versa if we choose one of the spins in one of the states of level $\underline{P}_j$ at
random and apply the mechanism to it, then the probability of arriving
at level $\underline{P}_i$ is $V_{ij} / (N \, \Omega(\underline{P}_j))$. 

This simple observation is used to create an algorithm which serves
two purposes:

i) It determines the rate of attempts $T_{ij} =
V_{ij}/(N \, \Omega(\underline{P}_i))$ to go from a starting level
$\underline{P}_i$ to another level $\underline{P}_j$ which can be
reached by the chosen mechanism.

ii) It leads to an equal probability of visiting the levels
irrespective of their degeneracy.

Once the $T_{ij}$ have been found it is an easy matter to construct
the whole surface $\Omega(\underline{P})$ from the ratios $T_{ij}/T_{ji} =
\Omega(\underline{P}_j)/\Omega(\underline{P}_i)$.

In order to calculate the $T_{ij}$ one writes down the total number
$Z_i$ of applications of the mechanism while in level
$\underline{P}_i$ and also the number of applications $Z_{ij}$ which
would lead to level $\underline{P}_j$ and this in both cases, if the step is executed or not. Then
$T_{ij} = Z_{ij}/Z_i.$

As a reversible transition mechanism we choose single spin flips. Before the
spin is flipped the system is characterized by the parameter values
$\underline{P}_i$. A spin is chosen at random. Its flip would lead to
a new state characterized by $\underline{P}_j$. We add 1 to $Z_i$ and
also to $Z_{ij}$ whether the flip is accepted or not. After a large
number of attemped spin flips $Z_{ij}/Z_i$ approches $T_{ij}$. The
demand ii) is fulfilled, if steps from $i$ to $j$ and from $j$ to $i$
are executed with the same frequency. This can be achieved by
introducing probabilities for the acceptance of executing a step. This
probability of accepting the step from $i$ to $j$ is equal to 1 if
$T_{ij} < T_{ji}$ and it is $T_{ji} / T_{ij}$ otherwise. This
equalizes the probabilities

\begin{equation} \label{Gl:6}
p(\underline{P}_i,\underline{P}_i + \Delta \, \underline{P}) =
p( \underline{P}_i + \Delta \, \underline{P}, \underline{P}_i)
\end{equation}
of going from a level $\underline{P}_i$ to the level $\underline{P}_i
+ \Delta \, \underline{P}$ and of the reverse transition.




Choosing nonzero, but otherwise arbitrary values for $Z_i$ and
$Z_{ij}$ for all $i$ and $j$ and starting from any one of the states
of the system one rapidly builds up good estimates for the transition
rates $T_{ij}$. When this point is reached all levels
$\underline{P}_i$ are visited with the same frequency and the quality
of the transition variables is improved with the same rate over the
whole parameter range.  

The procedure adopted here differs from other implementations\cite{Oli98,Olib98,Wan98,Wan00} of the
transition variable method in several respects: The transition
variables are updated at every single spin flip, in return we do not
register all the possible transitions which are possible from a given
microstate, but only those ones which are attempted by the algorithm,
whether they are successful or not.
Furthermore the transition probabilities $T_{ij}$ are updated
during the whole run and the acceptance rates are governed by the most
recent value of $T_{ij}$ from the beginning of the simulation to its
very end.

This method gives us the freedom of restricting the calculations to a chosen
number of channels in energy or magnetization direction. Whereas for not too
large systems all values of the energy and of the magnetization may be allowed
in a single run, for larger system sizes 
narrow bands in energy direction covering all magnetization channels may be used.
In the limit of only one energy channel per band our approach differs from the
well-known Q2R method \cite{Her86,Lan87,Sta91}, as for a fixed energy value all the possible magnetizations are
visited with the same frequency.

Our algorithm has been subjected to several stringent tests. We have
calculated $\Omega_N(E)$ for a 32 x 32 Ising model and we have
compared the resulting entropy with the exact data of Beale\cite{Bea96}. $S_N(E)$
as well as its first and second derivate match the exact values almost
perfectly. 
For example, $7 \times 10^5$ Monte Carlo sweeps yield for the inverse microcanonical
temperature $S'_N(E)$ an average error as small as $0.025 \%$ when compared to the first
derivative of the Beale data.
In a plot of the specific heat $c(E) = - (S'_N(E))^2 /
S''_N(E)$ versus temperature $T(E)$, where $1/T(E) = \beta^*(E) =
S'_N(E)$, one cannot distinguish the simultated data from the exact
result. For the same system we have also calculated
$\hat{\Omega}_N(E,M)$ from which we again find $\Omega_N(E)$, now by
summation over M. In
addition we find $\tilde{\Omega}_N(M)$ by summation over $E$ and there is
exellent agreement with the exact result which is $N!/[\frac{N+M}{2}!
\frac{N-M}{2}!]$.

The algorithm compares very favorably to other methods proposed for the computation
of the density of states. Its efficiency for the 32 x 32 Ising model is slightly better
than that of the multiple-range random
walk algorithm of Wang and Landau\cite{Wan01} where an average error of $0.035 \%$ was obtained
for $7 \times 10^5$ Monte Carlo updates. It must be stressed that our method is completely different 
to the Wang/Landau method where a rough estimator of the DOS is built up
rapidly due to a multiplicative update of $\Omega(E)$ every time
the energy level $E$ is visited. The error introduced by this
multiplicative update has to be eliminated before obtaining the
DOS with high precision \cite{Wan01,Hue01}. In our present approach, no such error
is introduced during the course of the simulation.

As we will show in the following, the new method enables us to compute the density
of states with a high accuracy even for large Ising systems. Furthermore,
preliminary results for complex systems with a rough energy landscape
are very promising \cite{Fro01}.

As a first application, we compute in the next Section the microcanonical order
parameter in the two- and three-dimensional Ising models. It must be noted that data
of extremely high accuracy, rarely needed in a canonical study, are mandatory for
a reliable determination of microcanonically defined quantities.

\section{Microcanonically defined spontaneous magnetization}
In a microcanonical analysis of finite systems with $N = L^d$ spins ($d$ being the number of space
dimensions), the object of interest is the density of states $\Omega_N(E,M)$ as a function of
the energy $E$ and the magnetization $M$. We define the spontaneous magnetization $M_{sp,N}(E)$
as the value of $M$ where the entropy $S_N(E,M) = \log \Omega_N(E,M)$ at a fixed
value of the energy has its maximum with respect to $M$.
At energies lower than a critical value $E_{c,N}$ the entropy exhibits two maxima at
$M = \pm M_{sp,N}$. On approach to this point $E_{c,N}$ the order parameter $M_{sp,N}$
vanishes with a square root behaviour in the finite system \cite{Kas00}.

In order to achieve very good statistics we have restricted
the calculations to a narrow band which covers only five
channels in energy direction, but stretches over all possible values of the
magnetization.
All single spin flips within this band are allowed. Spin flips
which would end up outside the band are rejected. We build up the
$T_{ij}$ and from them construct the DOS in magnetization
direction. A single flip leads from a microstate in the level $\underline{P}_i
= (E_i,M_i)$ with entropy $S_N(E_i,M_i)$ to a state of a neighbouring level
$\underline{P}_j = (E_j,M_j)$ with $E_j = E_i + 4 \, \eta$ and
$M_j=M_i + 2 \gamma$. Here $\gamma = \pm 1$, whereas $\eta = 0$, 
$\pm 1$, $\pm 2$ for a two-dimensional square lattice and
$\eta = 0$, $\pm 1$, $\pm 2$, $\pm 3$ for a three-dimensional simple cubic lattice. The entropy
at the macro state at $\underline{P}_j$ can then be expressed as:
\begin{eqnarray} \label{Gl:7}
S_N(E_j,M_j) = && S_N(E_i,M_i) + \beta^*(E_i,M_i)\, 4 \, \eta \, \nonumber \\
&& + \mu(E_i,M_i)\, 2 \, \gamma \, + \ldots
\end{eqnarray}          
where $\beta^*(E,M) = \partial S_N(E,M)/\partial E$ is the inverse
temperature and $\mu(E,M) = \partial S_N(E,M)/\partial M$. 
Because of the
densely spaced energy and magnetization levels (on the intensive scale)
higher order terms in (\ref{Gl:7}) may savely be neglected. 
From ratio
of one transition observable pair $T_{ij}/T_{ji} = \Omega(\underline{P}_j)/
\Omega(\underline{P}_i)= \exp \left[ 4 \eta \beta^* + 2 \gamma \mu \right]$ follows
one linear combination of $\beta^*$ and $\mu$. In two dimensions, the ten states
at the magnetizations $M$ and $M+2$ and energies $E_0 + 4 \eta$ within the
band of states to which the calculation was restricted, are connected
by 12 transition variable pairs. Averaging over the 12 equations which determine
$\beta^*$ and $\mu$ considerably increases the statistical relevance of the data.
In three dimensions, the number of transition variable pairs is even larger.

Thus, the transition variable method directly provides the derivative $\mu(E,M)$
of the entropy which we need for the determination of its maximum.

From the zeros of $\mu(E,M)$ the spontaneous magnetization is
determined with high precision as shown in Figure 1. 
The variation of the spontaneous magnetization 
per spin, $m_{sp,N}=M_{sp,N}/N$, as a function of
the specific energy, $\varepsilon = E/N$, is displayed in Figure 2 
for two-dimensional Ising models on an infinite
and on a small square lattice with $32^2$ spins.
The curve for the infinite lattice has been drawn from the exact results for
$M(T)$ and $E(T)$ \cite{McC73}.
Both curves coincide at low energies.
In our analysis,
$M_{sp}(E)$ for energies lower than the critical value $E_{c,\infty}$ of the
infinite system is fitted by

\begin{equation} \label{Gl:8}
M_{sp}(E) = A \left| \varepsilon^* \right|^{\beta_{\varepsilon,eff} (E)}       
\end{equation}
with an energy dependent effective exponent
$\beta_{\varepsilon,eff}(E)$. Here, $\varepsilon^* = (E_{c,\infty}-E)/(E_{c,\infty}-E_g)$ is the reduced energy.
$E_g$ denotes the ground state energy.
The effective exponent, which is given by $\beta_{\varepsilon,eff}=\mbox{d} \ln M_{sp}/\mbox{d} \ln \left|
\varepsilon^* \right|$, yields the critical exponent $\beta_\varepsilon$ in the limit $E \longrightarrow
E_{c,\infty}$.
Contrary to Ref.\cite{Kas00}  the critical expansion refers
to $E_{c,\infty}$ and not to the critical energy of the finite
system. The procedure has been performed for a series of system
sizes $N_1<N_2<N_3 \ldots$ In the energy range where $N_i$ and
$N_{i+1}$ yield the same exponent $\beta_{\varepsilon,eff}(E)$ this exponent is
selected for the plot in Figure 3. This value of $\beta_{\varepsilon,eff} (E)$ is
used up to the energy where the two exponents start to differ from
each other, i.e.\ until finite size effects show up. 
Then we plot the common exponent for the system sizes
$N_{i+1}$ and $N_{i+2}$ until these begin to disagree
and so on. The
largest systems for which we have performed these calculations were
$700 \times 700$ in two dimensions and $80 \times 80 \times 80$ in three
(these are presumably the largest Ising systems for which microcanonical quantities
have ever been determined). 
This approach has the advantage that the finite-size effects are monitored
closely. Furthermore, one avoids to simulate unnecessarily large 
systems at low energy where small systems already yield data essentially
free of finite-size effects.

\begin{figure}
\centerline{\psfig{figure=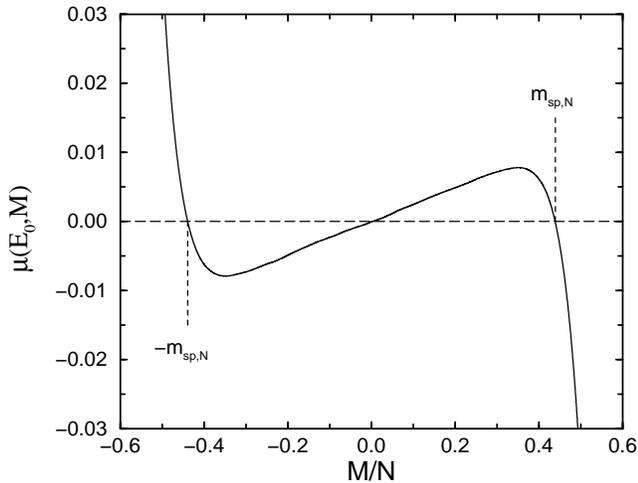,angle=270,width=3.3in}}
\caption{Derivative $\mu(E_0,M) = \partial S(E_0,M)/\partial M$ of the entropy of
a three-dimensional Ising model with $20 \times 20 \times 20$ spins computed at
the energy $E_0/N=-1.2$. The spontaneous magnetization $M_{sp,N}/N = \pm 0.441$ is obtained from
$\mu(E_0,M)=0$. For energies $E < E_{c,N}$, a local minimum of the entropy is observed 
at $M=0$.}
\end{figure}

The critical exponent $\beta_\varepsilon$ is related to the commonly used
exponent $\beta$ which describes the behavior of the order parameter
with respect to the temperature by $\beta_\varepsilon =
\beta/(1-\alpha)$ where $\alpha$ is the usual specific heat 
critical exponent \cite{Kas00}. In two dimensions where $\alpha = 0$ we expect
$\beta_\varepsilon = 1/8$, in three dimensions with $\beta = 0.327$ and
$\alpha = 0.11$ we expect $\beta_\varepsilon = 0.367$. 

Also shown in Figure 3 is the corresponding effective exponent derived
from the exact result $M(E)$ of the infinite two-dimensional Ising model\cite{McC73}.
Obviously, the numerical data follow closely the exact result, thus
demonstrating the reliabilty of our algorithm for computing accurate
microcanonical quantities even for large systems.

\begin{figure}
\centerline{\psfig{figure=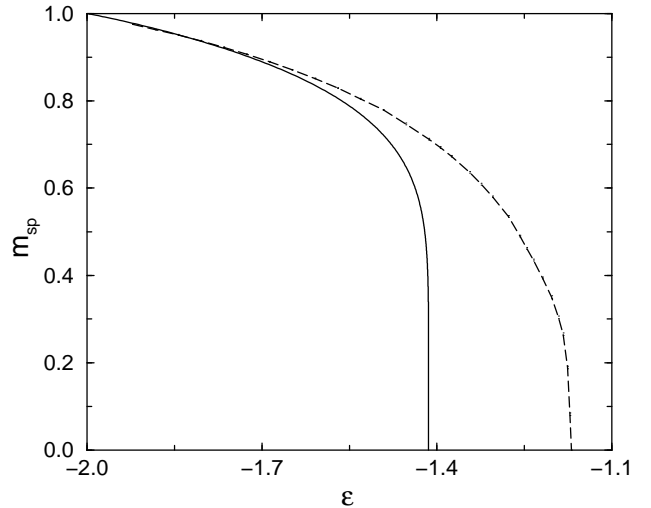,angle=270,width=3.3in}}
\caption{
Microcanonically defined order parameter vs specific energy for two-dimensional Ising models
defined on an infinite square lattice (full line) and on a finite square lattice containing
$32 \times 32$ spins (dashed line). The finite-size spontaneous magnetization vanishes at a well-defined
finite-size critical point.}
\end{figure}

\begin{figure}
\centerline{\psfig{figure=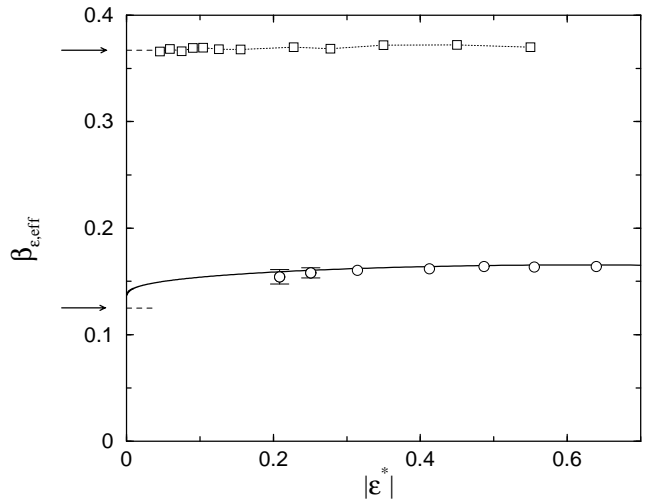,angle=270,width=3.3in}}
\caption{Effective exponent $\beta_{\varepsilon,eff}$ as function of the reduced energy $\left|
\varepsilon^* \right|$ obtained for the Ising model in two (circles) and three (squares) 
dimensions. Only error bars larger than the symbol sizes are shown. The dashed lines 
and the arrows indicate the expected
values of the order parameter critical exponent $\beta_\varepsilon$ in two and three 
dimensions. The full line is the effective exponent derived from the exact result
$M(E)$. Note its rapid decrease towards the value $1/8$ very close to the critical point.}
\end{figure}  
  

Figure 3 exhibits two pecularities of the 2D system: (i) The strong $N$ dependence
of $E_{c,N}$ makes it extremely difficult to obtain data points for $\left| \varepsilon^*
\right| < 0.2$. (ii) The effective critical exponent $\alpha_{eff} (t)$ of the
specific heat $c(t)$ of the infinite system is finite for finite $t=(T_c-T)/T_c$,
but precipitates to zero as $t=0$ is approached. This causes the dramatic behavior
of $\beta_{\varepsilon,eff}(\varepsilon^*)=\beta_{eff}/(1-\alpha_{eff})$ in the vicinity
of $\left| \varepsilon^* \right| =0$.

Remarkably, good values of the exponents are already found for the
smallest systems considered here, namely $L = 50$ in two and $L = 10$ in three
dimensions. This is especially true for the three-dimensional case where,
over the whole energy range considered,
the value of the effective exponent does not noticeably differ from the
value of the critical exponent. Even for energies more than 30 \% below 
the critical energy and for systems consisting of only 1000 spins, the
obtained effective exponent is almost identical to the critical exponent.
This is in marked contrast to a similar canonical analysis of the
behaviour of the order parameter with respect to the temperature where large
corrections to scaling lead to an effective temperature dependent exponent 
which differs by almost 50 \% from the value of the critical exponent at temperatures
some 30 \% below $T_c$\cite{Ple98}. Whereas it is obvious from Figure 3 that corrections are
present when analysing the two-dimensional Ising model, these corrections
are again small compared to the corrections observed for the temperature dependent
effective exponent derived from the order parameter $M(T)$. 

The origin of the different behaviour of $\beta_{\varepsilon,eff}$ as function
of the reduced energy in two and three dimensions can be traced back to the
logarithmic
divergence of the specific heat in the two-dimensional Ising model.
We expect a behaviour similar to that observed in the three-dimensional Ising
model for all models with a non-vanishing specific heat critical exponent.
Future investigations of various models will be needed to clarify this point.


\section{Summary}
We have developed an algorithm which rapidly builds up transition
variables for any reversible rule of jumping from one macrostate of a
system, characterized by a set $\underline{P}$ of parameters, to another one. The most
recent values of the transition variables govern the further path of the
system through phase space in such a way that equally good statistics is
obtained over the whole area which is covered by the calculation. As an
example we have chosen the two- and three-dimensional Ising models with $\underline{P}=(E,M)$
together with single spin flips as the transition mechanism. From the
transition observables we obtain highly accurate data for $dS/dM$ which
allow a precise determination of the microcanonically defined
spontaneous magnetization $M_{sp}(E)$. An analysis of $M_{sp}(E)$ for
different system sizes yields 
excellent values for the order parameter critical exponent
$\beta_{\varepsilon}$.
 

\end{document}